\documentclass[runningheads]{llncs}
\usepackage[T1]{fontenc}

\usepackage{silence}
\usepackage{subcaption}
\usepackage{pifont}
\usepackage{array} 
\usepackage{tabularx} 
\usepackage{minted} 
\usepackage{xurl}
\usepackage{multicol}
\usepackage{multirow}
\usepackage{makecell} 

\usepackage{balance}
\usepackage{float}
\usepackage{textcomp}
\usepackage{cite}

\usepackage{adjustbox}

\usepackage{enumitem}
\usepackage{booktabs}

\usepackage{tikz}
\usetikzlibrary{arrows.meta, positioning, shapes.multipart, fit,calc}
\usepackage{amsmath,amssymb,amsfonts}
\usepackage[colorlinks=true,urlcolor=black]{hyperref}
\usepackage[capitalise,noabbrev]{cleveref}

\usepackage{algorithmic}
\usepackage{graphicx}
\usepackage{textcomp}
\usepackage{bmpsize}
\usepackage{xcolor}
\usepackage{lipsum}
\def\BibTeX{{\rm B\kern-.05em{\sc i\kern-.025em b}\kern-.08em
    T\kern-.1667em\lower.7ex\hbox{E}\kern-.125emX}}

\usepackage{listings}
\usepackage{xcolor}

\usepackage{tabularx}
\newcolumntype{L}{>{\raggedright\arraybackslash}X}

\usepackage{xurl}
\usepackage[final]{microtype}
\begin{document}

\title{To WASM or Not to WASM: Evaluation of Browser Fingerprinting Defenses Under WASM based Obfuscation}

\titlerunning{To WASM or Not to WASM}

\author{
A H M Nazmus Sakib\inst{1}\thanks{Corresponding author} \and
Mahsin Bin Akram\inst{1} \and
Joseph Spracklen\inst{1} \and
Sahan Kalutarage\inst{1} \and
Raveen Wijewickrama\inst{1} \and
Igor Bilogrevic\inst{2} \and
Murtuza Jadliwala\inst{1}
}

\authorrunning{Sakib et al.}

\institute{
The University of Texas at San Antonio
\email{\{ahmnazmus.sakib, mahsin.akram, joseph.spracklen, sahan.kalutarage, raveen.wijewickrama\}@my.utsa.edu, murtuza.jadliwala@utsa.edu}
\and
Google LLC
\email{ibilogrevic@google.com}
}

\maketitle

\begin{abstract}
Browser fingerprinting defenses have historically focused on detecting JavaScript(JS)-based tracking techniques. However, the widespread adoption of WebAssembly (WASM) introduces a potential blind spot, as adversaries can convert JS to WASM's low-level binary format to obfuscate malicious logic. This paper presents the first systematic evaluation of how such WASM-based obfuscation impacts the robustness of modern fingerprinting defenses. 
We develop an automated pipeline that translates real-world JS fingerprinting scripts into functional WASM-obfuscated variants and test them against two classes of defenses: state-of-the-art detectors in the research literature and commercial, in-browser tools. 
Our findings reveal a notable divergence: detectors proposed in the research literature that rely on feature-based analysis of source code show moderate vulnerability, stemming from outdated datasets or a lack of WASM compatibility.
In contrast, defenses such as browser extensions and native browser features remained completely effective, as their API-level interception is agnostic to the script's underlying implementation. These results highlight a gap between detection approaches proposed in prior literature and the practical defenses deployed in modern browsers, offering insight into how detection methods can be strengthened against WASM-based obfuscation while also indicating how attackers may develop more evasive techniques.
\keywords{WASM \and Web Assembly \and Fingerprinting \and Obfuscation}
\end{abstract}

\section{Introduction}
\label{sec:introduction}

Browser fingerprinting enables websites and third-party trackers to uniquely identify users by collecting and analyzing subtle differences in device and browser behavior. These differences arise from variations in canvas rendering, WebGL capabilities, audio processing, hardware features, and other low-level properties~\cite{laperdrix2016beauty, acar2014web}. 
Recent studies have shown that approximately 69\% of the top 10,000 websites from the Alexa 100\textit{K} list\cite{alexa100k} employ some form of fingerprinting, often without user awareness or consent~\cite{fp_inspector_paper}.
The browser fingerprinting market itself is projected to grow significantly, from a valuation of \$2.5 billion in 2023 to over \$12.5 billion by 2031, with a compound annual growth rate (CAGR) of 25\%~\cite{browser_market_growth}. This growth underscores the increasing reliance on fingerprinting for analytics, advertising, and surveillance.

Because these fingerprinting techniques do not rely on traditional identifiers such as cookies or IP addresses, they operate covertly and passively, often making them difficult to detect or block~\cite{englehardt2016online}.
In response, researchers and browser vendors have proposed detection tools and defenses that largely focus on JavaScript(JS)-based fingerprinting. This is a natural focus, as JS remains the dominant language for client-side execution and is commonly used to extract device- and browser-level attributes~\cite{nikiforakis2013cookieless, englehardt2016online, acar2013fpdetective}.

However, this JS-centric focus overlooks recent changes in the web platform, most notably the growing adoption of WebAssembly (WASM)~\cite{haas2017bringing}. 
WASM is a low-level, portable binary instruction format designed to complement JS, offering performance close to native code. As of 2024, WASM was supported in more than 96\% of browsers worldwide~\cite{wasm_usage}. Beyond legitimate performance-critical applications, WASM has also been adopted by adversaries as a tool to obfuscate JS code and hide malicious functionality~\cite{wobfuscator}.
A compelling example is Wobfuscator~\cite{wobfuscator}, which demonstrated that selectively translating JS into WASM can evade state-of-the-art \emph{static} malware detectors; the authors explicitly scoped their evaluation to static analysis of general malware and conceded that their technique would likely fail against dynamic detectors that observe runtime API calls. In parallel, several recent works have used WASM alongside JS to implement new fingerprinting techniques~\cite{trampert2022, pisano2023, botvinnik2023}, and Guri et al.~\cite{guri2025} showed that fingerprinting logic can be implemented natively in WASM without relying on JS, expanding the adversarial surface.

While prior work hypothesized that dynamic defenses might catch WASM obfuscation \cite{wobfuscator}, this claim remains empirically untested for API-heavy threats. Browser fingerprinting represents a fundamentally different threat model than general malware, relying predominantly on continuous invocations of JS-exposed rendering and sensory APIs. Unlike standalone malicious payloads that can potentially execute completely in isolation, fingerprinting scripts are almost inextricably linked to these observable, runtime probes. With many fingerprinting defense research incorporating dynamic scanners\cite{fp_inspector_paper,boussaha24fptracer}, a transition of fingerprinting techniques to WASM exposes a critical, unexplored security gap. This raises the following question:
\textbf{\emph{Can adversaries leverage WASM to obfuscate browser fingerprinting in ways that evade both static and dynamic defenses, including those used in research and real-world deployments?}}
Specifically, is it feasible to convert JS-based scripts into WASM in a way that preserves functionality while bypassing these disparate systems? Our goal in this paper is to translate known JS fingerprinting techniques into WASM to empirically assess the true robustness of existing fingerprint detection systems. To this end, we advance the adversarial model from the selective, \emph{opportunistic} translation seen in prior work \cite{wobfuscator}  to a highly aggressive \emph{greedy translation strategy}. Here, the conversion pipeline attempts to translate as much JS code as possible into WASM without filtering or selective heuristics. We hypothesize that maximizing the scope of translation, rather than masking isolated components, better obscures the syntactic and semantic patterns used by fingerprinting detectors. Since many detection systems operate as black boxes, this greedy strategy enables a comprehensive stress test across diverse techniques under aggressive WASM conversion.

Furthermore, we convert a large corpus of real-world JavaScript into WebAssembly to assess the viability of this attack vector at scale. This large-scale conversion allows us to (1) evaluate the execution viability and functional equivalence of the JS-to-WASM transformation pipeline, (2) ensure that our analysis covers diverse, real-world script behaviors, and (3) characterize the structural changes introduced by WASM obfuscation that may inform future countermeasures.
In summary, this paper makes the following key contributions:
\begin{itemize}[leftmargin=*, itemsep=-0pt]
  \item \textbf{Conversion Pipeline.} To explore whether real-world JS can feasibly be converted to WASM as part of an automated obfuscation pipeline, we build a conversion framework which is capable of processing a wide range of real-world fingerprinting scripts, maintaining functionality while potentially obscuring key logic.

  \item \textbf{Datasets.} We build two complementary, first-of-their-kind datasets to enable comprehensive evaluation of WASM-based fingerprinting evasion: (i) a large-scale corpus of 7.5 million JS scripts (with over 10\textit{K} fingerprinting scripts) and their WASM-obfuscated counterparts using our conversion framework, and (ii) a controlled dataset of 124 fingerprinting scripts, each with paired JS and WASM (converted) versions spanning multiple representative fingerprinting categories.
  \item \textbf{At-Scale Measurement of the Deployed Defense Ecosystem.} We conduct the first large-scale empirical measurement of how JS-to-WASM obfuscation affects modern fingerprinting defenses, spanning two SOTA detectors from research literature, ten anti-fingerprinting browser extensions, five browser configurations, and one WASM-native fingerprinting technique. We additionally implement a dynamic analysis–based fingerprinting detector to evaluate whether runtime behavioral signals remain robust under WASM obfuscation.
  \item \textbf{Root Cause Evaluation.} We analyze the detection outcomes of WASM-converted scripts to uncover the factors underlying successful or failed evasion of existing fingerprinting detectors.
\end{itemize}

Our experiments reveal a clear disparity: academic detectors relying on JS source-code analysis are considerably weakened by WASM obfuscation (recall drops up to 33 percentage points), while commercial browser extensions and built-in browser protections remain fully effective because their API-level interception is language-agnostic. A dynamic, API-level monitor we implement on a modern stack confirms this robustness. We also evaluate the only known native WASM fingerprinting technique~\cite{guri2025} and observe that it bypasses all tested defenses except those that block WASM entirely.
It should be noted that we use the terms \emph{WASM conversion} and \emph{obfuscation} interchangeably when discussing the transformation of JS into WASM. 

\section{Background and Related Work}
\label{sec:background-related}

\label{sec:background}

\noindent
\textbf{Browser Fingerprinting.}
Browser fingerprinting is a stateless tracking method that relies on collecting distinctive attributes from a user's browser and device configuration. These attributes include JS-exposed APIs (e.g., WebGL, canvas), HTTP headers (e.g., User-Agent), browser plugins, system fonts, screen resolution, and more~\cite{laperdrix2016beauty, acar2014web, eck}. Unlike first-party or direct tracking methods such as cookies, fingerprinting does not store data on the client and can operate without user awareness or consent~\cite{englehardt2016online}. 

The concept of fingerprinting was first formalized in 2010 by Eckersley~\cite{eck} following Mayer’s earlier observations in 2009~\cite{mayer}. Subsequent work expanded fingerprinting techniques to include NoScript behavior~\cite{mowery}, JS engine quirks~\cite{mulazzani}, CSS rendering features~\cite{css1_1, css1_2}, and canvas/WebGL rendering inconsistencies~\cite{canvas_1, acar2014web}. Researchers have also exploited hardware-level behavior~\cite{hd1, hd2}, audio APIs \cite{englehardt2016online}, screen glyphs~\cite{font_glyphs}, and even battery APIs~\cite{battery} as fingerprinting vectors.

\noindent
\textbf{Defenses Against Fingerprinting.}
Researchers and developers have proposed various defenses aimed at detecting or disrupting fingerprinting behavior.
Defensive strategies against fingerprinting fall into three main categories:

\emph{(1) API Randomization and Normalization.}  
This category includes techniques that modify the outputs of fingerprint-relevant APIs to either randomize them across sessions or normalize them across users. The goal is to reduce the uniqueness or stability of the collected attributes, thereby degrading fingerprinting accuracy. Tools such as Canvas Defender~\cite{c_def} and Firegloves~\cite{firegloves} inject noise into rendering APIs, though such noise can itself become a fingerprint~\cite{random_to_fp}. More principled normalization strategies are used in browsers like Brave and Tor~\cite{uni_brave,uni_tor}, but they could potentially break site functionality~\cite{fp_inspector_paper}.

\emph{(2) Heuristics-Based Detection.}  
Heuristic defenses attempt to identify fingerprinting behavior using manually curated rules or behavioral patterns. These tools flag scripts that use certain APIs or access a suspicious number of device properties. Some examples of such tools include Privacy Badger~\cite{privacy_badger} and Disconnect~\cite{disconnect}, which operate at the browser extension level. However, their effectiveness depends heavily on timely rule updates and may not generalize well to novel fingerprinting strategies.

\emph{(3) Learning-Based Detection.}  
This category includes machine learning (ML) approaches that use static, dynamic, or hybrid features extracted from JS code to classify scripts as fingerprinting or benign. Tools such as FP-Radar~\cite{fp_radar} and FProbe~\cite{fprobe} analyze web API usage over time or via control/data flow analysis. DeepFPD~\cite{deepfpd} uses token sequences, abstract syntax trees (AST), and control flow graphs (CFG) as input features. FP-Inspector~\cite{fp_inspector_paper} combines both static and dynamic signals using ML, and has been evaluated on a large-scale corpus of real-world websites~\cite{alexa100k}.

\noindent
\textbf{WebAssembly (WASM).}
WASM is a low-level, portable binary instruction format designed to complement JS by enabling high-performance execution in web browsers ~\cite{wasm_1}. It is widely used for performance-intensive purposes 
such as gaming, multimedia processing, and runtime environments~\cite{wasm_4}.
 Unlike JS, WASM runs closer to native hardware and lacks a high-level structure, making it more difficult to inspect and analyze statically. These properties, while beneficial for performance, also make WASM attractive for adversarial use cases, including 
 code obfuscation and evasion of security mechanisms~\cite{wobfuscator}.

Prior fingerprinting research~\cite{fp_inspector_paper,deepfpd,fp_radar,fprobe,eck,mayer,mulazzani,canvas_1,acar2014web} has focused primarily on JS-based attacks and defenses. However, the broad adoption of WASM introduces new challenges: its binary format and limited semantic cues make conventional 
static/dynamic 
analysis challenging for detecting WASM-implemented behavior.

Trampert et al.~\cite{trampert2022} and Pisanò et al.~\cite{pisano2023} leveraged WASM for performance and stability in fingerprinting routines, such as CPU and audio-based tests. Botvinnik et al.~\cite{botvinnik2023} used WASM for power-based fingerprinting due to its portability. Guri et al.~\cite{guri2025} proposed a WASM-native fingerprinting approach using timing variations caused by interactions between WASM and JS to differentiate browsers. However, none of these studies evaluated whether such techniques can evade existing fingerprinting detectors.

A separate line of work uses WASM as an \emph{obfuscation} layer. Romano et al.~\cite{wobfuscator} introduced Wobfuscator, which demonstrates that JS code can be obfuscated through selective translation into WASM, helping general malware evade static detectors. Their approach, termed \textit{opportunistic translation}, preserves functionality while obscuring structure but, as they observe, is not expected to defeat dynamic detectors that observe runtime API invocations. Expanding on this adversarial surface, Cabrera-Arteaga et al. \cite{CABRERAARTEAGA2023103296} applied WASM diversification to further enhance malware evasion, while Xia et al. \cite{xia_malware} proposed static semantics reconstruction to improve the detection of these hybrid JS-WASM payloads. However, these works focus strictly on malwares, not fingerprinting.

To date, no comprehensive study has examined how obfuscating JS-based fingerprinting techniques using WASM affects their detectability across modern detection tools, nor has any prior work in the literature systematically evaluated native WASM-based fingerprinting in this context. Our work addresses this gap by applying WASM-based obfuscation to real-world fingerprinting scripts, assessing their detectability across a representative cross-section of academic and deployed defenses, and analyzing the underlying causes of detection failures.

\section{Adversary Model and Research Questions}
\label{sec:adv_model_rqs}

\begin{figure}[ht]
    \vspace{-0.3cm}
    \centering
    \includegraphics[width=1\linewidth]{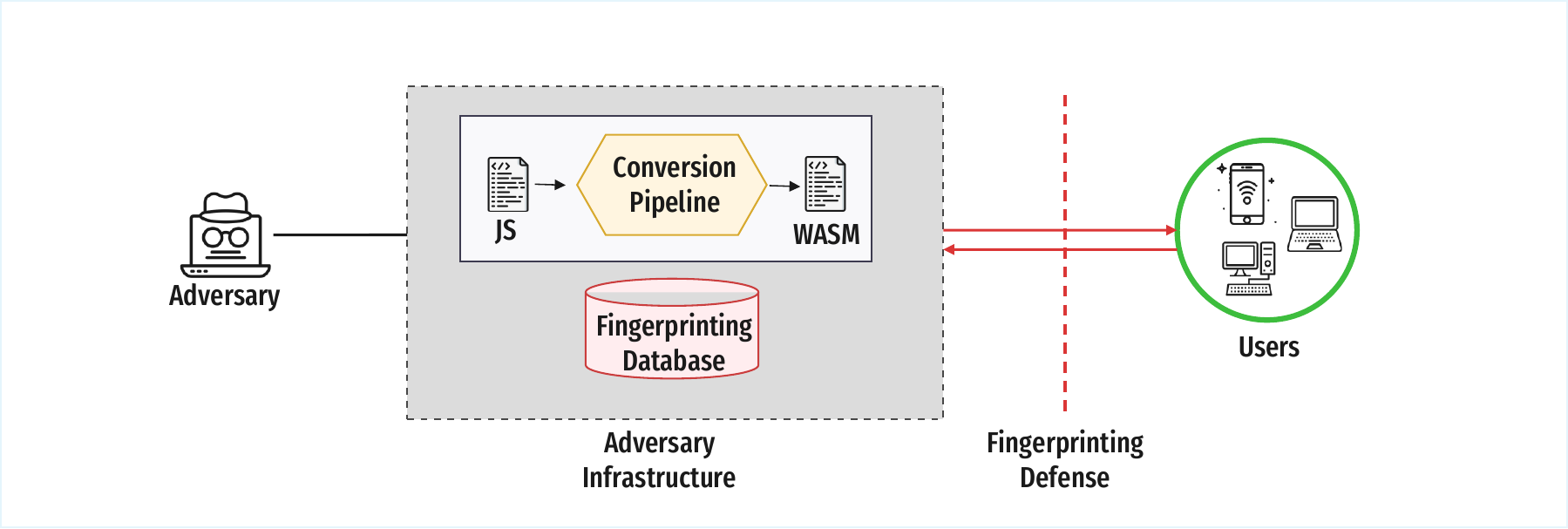}
    \vspace{-0.7cm}
    \caption{Adversary model.}
    \label{fig:Adversary Model}
\end{figure}

\subsection{Adversary Model}
\label{sec:adv_model}

Our adversary is a web‐based tracker whose goal is to harvest the same rich set of device and browser attributes (e.g., canvas fingerprints, WebGL parameters, audio-subsystem characteristics) that traditional JS-based fingerprinting tools extract, but without triggering existing detection mechanisms (see \cref{fig:Adversary Model}). The adversary controls the server or third-party domain delivering client-side code and can freely transform scripts prior to delivery.
The adversary can (i)~translate JS fingerprinting scripts into WASM to obfuscate their implementation while preserving functionality, and (ii)~implement fingerprinting logic natively in WASM (relying on lower-level interactions rather than JS-based probe).

In the scenario where the adversary converts JS scripts into WASM, we assume a \textit{greedy conversion strategy}, where the adversary attempts to translate as much JS code as possible into WASM rather than selectively targeting specific functions. This maximizes code coverage within each script and increases the chance of evading syntactic or semantic detection heuristics. The one-time conversion overhead is easily amortized, as fingerprinting scripts are reused across many users.
We assume an off-the-shelf, up-to-date browser (e.g., Chrome or Firefox) with no special instrumentation or privileged access, reflecting the typical environment encountered by real-world fingerprinters. The adversary cannot escape the JS/WASM sandbox, or tamper with the browser engine, as such capabilities extend beyond fingerprinting and would render fingerprinting unnecessary.
From a defense perspective, we assume that the user may employ 
both static and dynamic fingerprinting detectors (i.e., tools that analyze code statically without execution or dynamically based on runtime behavior). We also assume that the adversary has no knowledge of the defense model’s architecture, parameters, or training data, if the defense employs a machine learning-based approach.

\subsection{Research Questions}
\label{sec:rqs}

To assess the feasibility and detection implications of WASM-based browser fingerprinting, we structure our investigation around the following three RQs.

\begin{itemize}[leftmargin=*, itemsep=-0pt]
    \item \textbf{RQ1 (Conversion Feasibility):} 
    
    Can existing JS fingerprinting scripts be reliably translated into WASM while preserving functional behavior across diverse fingerprinting categories? We measure success rates, conversion coverage, time, and size overhead at scale, plus stable-hash output equivalence on a controlled dataset.
    
    \item \textbf{RQ2 (Detection Robustness):} 
    How well do the SOTA fingerprinting detection tools perform when faced with WASM-based implementations? Specifically, we analyze the effectiveness of detectors proposed in the research literature, browser-integrated defenses, and privacy extensions in detecting fingerprinting scripts that are either translated into or natively implemented in WASM.
    
    \item \textbf{RQ3 (Evasion Analysis):} 
    What structural, functional, or behavioral properties (of the scripts and/or detectors) contribute to detection failures or successes, and do current detection mechanisms exhibit blind spots or assumptions that can be exploited by adversaries?

\end{itemize}

\section{Methodology}
\label{sec:methodology}

To answer the research questions in \cref{sec:rqs}, we build a conversion pipeline that transforms JS-based fingerprinting scripts into semantically equivalent WASM-obfuscated variants and evaluate it through four experiments. We study scalability (RQ1), the impact on detectors and defenses (RQ2), and which transformations contribute most to evasion (RQ3).

\subsection{Conversion Pipeline}
\label{sec:cnv_ppl}

\begin{figure*}[htb]
    \vspace{-0.7cm}
    \centering
    \includegraphics[width=0.99\textwidth]{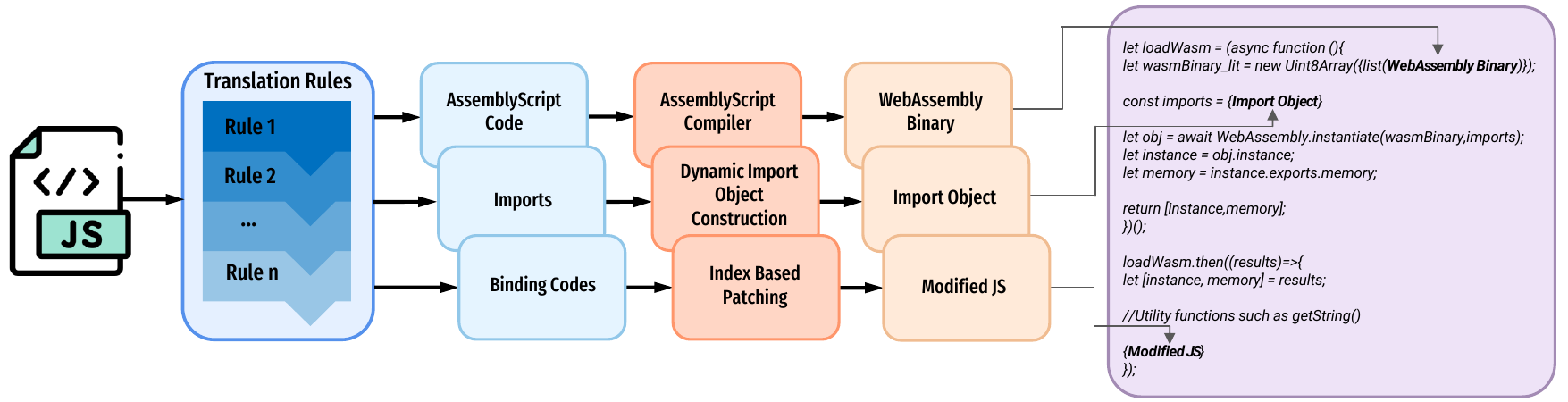}
    \vspace{-0.3cm}
    \caption{Conversion pipeline.}
    \label{fig:conversion_pipeline}
\end{figure*}

Converting JS to WASM is non-trivial because WASM is statically typed and stack-based, while JS is dynamically typed and event-driven. Many idiomatic JS features (e.g., closures, \texttt{eval}, async callbacks) do not map directly to WASM semantics, and browser execution still requires a thin JS wrapper to instantiate modules and mediate DOM access.

Despite these challenges, we deliberately embrace WASM translation to evaluate its impact on fingerprint detection. Under a black-box adversarial model (no knowledge of detector internals), it is unclear which script fragments drive classification; we therefore adopt a \textit{greedy conversion strategy} and translate as much JS as possible to maximize the WASM footprint and disrupt potential detection signals.
To this end, we design thirteen conversion rules: nine for general language constructs (e.g., literals, control flow, arrays, functions) and four for fingerprinting-relevant patterns (e.g., property access, dynamic element creation). Each rule replaces matched JS segments with semantically equivalent WASM or WASM-obfuscated logic. Several rules are inspired by \emph{Wobfuscator}~\cite{wobfuscator}, which performs \emph{opportunistic translation} by instantiating modules in place.
In contrast, our approach compiles all transformed regions into a single shared WASM module, which avoids initialization-order/timing issues from multiple asynchronous instantiations, simplifies integration, and reduces overhead.

We implement these WASM modules using \emph{AssemblyScript}~\cite{assemblyscript}, a statically typed subset of TypeScript that compiles to WASM. AssemblyScript is developer-friendly and closely mirrors JavaScript syntax, allowing us to write lightweight and high-performance WASM logic without needing to drop to low-level languages such as Rust or C/C++.

Before detailing the rules, we outline the structure of a converted script. The WASM startup is wrapped in an \textit{immediately invoked asynchronous function expression (IIAFE)} that embeds the .wasm bytes as a \texttt{Uint8Array}, supplies the import object, and calls \texttt{WebAssembly.instantiate()}. Once instantiation completes, the rewritten JS uses the exports/memory (via small helpers such as \texttt{getString} for UTF-8 decoding) and invokes the new WASM exports in place of the original JS logic. Following MDN Web Docs~\cite{mdn_webassembly}, we avoid blocking synchronous instantiation and rely on asynchronous \texttt{WebAssembly.instantiate()}.

The overall conversion pipeline (\cref{fig:conversion_pipeline}) is as follows:
\begin{enumerate}[leftmargin=*, itemsep=-0pt]
    \item The input JS code is parsed into an \textit{Abstract Syntax Tree (AST)}, which is recursively traversed to identify code fragments matching the syntactic patterns of the defined translation rules. We use the open-source Python library \emph{pyesprima}~\cite{pyesprima} to generate and manipulate the AST.

    \item For each matched fragment, the corresponding rule generates three artifacts: (i) an AssemblyScript snippet to be compiled into WASM, (ii) any required entries for the import object, and (iii) a JS glue snippet to interface with the compiled WASM module in place of the original code.

    \item The artifacts from all matched rules are aggregated: AssemblyScript snippets are merged into a single file and compiled into a WASM binary; import object entries are unified into a single import object; and glue code snippets are inserted into the original JS at their respective positions. Overlapping replacements are filtered out to prevent conflicts.
\end{enumerate}

\subsection{WASM Conversion Rules}
\label{subsec:rules-main}

\begin{table*}[t]
\vspace{-0.5cm}
\centering
\caption{Summary of the conversion rules.}
\label{tab:rule-summary}
\begin{adjustbox}{max width=\linewidth}
\begin{tabular}{@{}lll@{}}
\toprule
\textbf{Rule \#} & \textbf{Target JS Pattern} & \textbf{Purpose} \\
\midrule
1 & Literal-initialized variables & Offload literals to WASM and enable memory-based access \\
2 & Sensitive function calls (e.g., \texttt{eval}, \texttt{Function}) & Obfuscate sensitive invocations through indirect reconstruction \\
3 & Arrays of integers/floats & Move array data to linear memory to obscure structure \\
4 & Simple \texttt{if-else} statements & Offload control flow into WASM while preserving logic \\
5 & \texttt{for} loops & Obfuscate iteration logic by offloading loop structure \\
6 & \texttt{while} loops & Enable indirect evaluation of loop condition and body \\
7 & Function calls without return values & Wrap and obscure standalone invocations via WASM \\
8 & Class definitions & Obfuscate class declarations via string storage in WASM and \texttt{eval} \\
9 & Function declarations/expressions & Translate logic-heavy functions using LLMs to AssemblyScript \\
10 & Fingerprinting-relevant member expressions & Obfuscate access to high-signal APIs (e.g., \texttt{canvas.toDataURL}) \\
11 & Dynamic code generation calls & Convert runtime-generated expressions to indirect eval \\
12 & Fingerprinting relevant code inside strings (e.g., ``canvas'') & Mask known fingerprinting substrings via split reconstruction \\
13 & Hex-encoded property accesses using \texttt{screen} & Detect and obfuscate fingerprinting via encoded identifiers \\
\bottomrule
\end{tabular}
\end{adjustbox}
\vspace{-0.5cm}
\end{table*}

We define 13 transformation rules to convert JS code into WASM-compatible forms (summarized in \cref{tab:rule-summary}). Nine rules target general language constructs (e.g., literals, control flow, function declarations), and four focus specifically on browser fingerprinting behaviors. 
Inspired by Wobfuscator~\cite{wobfuscator}, our pipeline differs along four axes. Rule provenance: Rules~1--9 follow Wobfuscator’s construct categories, while Rules~10--13 are new and fingerprinting-specific. (i) \emph{Module structure.} Wobfuscator emits one WASM module per fragment; we instead use a single shared module with a \emph{greedy} strategy, avoiding initialization-order and async-instantiation issues from many modules. (ii) \emph{Code-shape coverage.} Per-fragment translation struggles with async callbacks, nested closures, and dynamic property access common in fingerprinting scripts; our shared-module design and Rule~9 explicitly handle these. (iii) \emph{Obfuscation goal.} Wobfuscator aims to do opportunistic translation, limiting their scope by design; whereas we target \emph{greedy conversion}: preserving execution while reshaping syntactic and string-level signals to the highest extent possible. (iv) \emph{Threat scope.} Wobfuscator evaluates against general-malware \emph{static} detectors and notes limitations against dynamic ones; we study whether WASM obfuscation weakens \emph{fingerprinting} defenses, including both static and runtime/API-level approaches.

Below, we highlight a subset of representative rules to demonstrate the breadth and novelty of our transformation pipeline. The details of the remaining rules are provided in~\cref{sec:remaining_rules}.

\noindent\textbf{Rule 1 – Replace Literals:} This rule replaces variable declarations initialized with literal values such as strings, integers, floats, or booleans with equivalent AssemblyScript exports. For example, a JS declaration such as:
\begin{lstlisting}
let x = 42;
\end{lstlisting}
is transformed into AssemblyScript:
\begin{lstlisting}
export let x_pos: i32 = 42;
\end{lstlisting}
To access the exported value at runtime, the following JS glue code is inserted:
\begin{lstlisting}
let x = instance.exports.x_pos;
\end{lstlisting}
For string literals, a helper function is used to reconstruct it from WASM linear memory by reading its length and decoding the corresponding byte sequence starting at a given pointer.
Literals inside loop headers (e.g., for-loop initializers or conditions) are excluded to preserve control flow behavior. All replacements are indexed to support precise patching into the original source.

\noindent\textbf{Rule 5 - Replace For Loops:} This rule targets ``for" loop constructs and replaces them with equivalent WASM-compatible representations.
The initializer (loop variable and its initial value), the loop condition, increment expression, and loop body are converted into AssemblyScript logic encapsulated within an exported function. Typical AssemblyScript code looks like:
\begin{lstlisting}
@external("js", "body_pos")
declare function body_pos(): void;

export function for_pos(): void {
let i: i32 = 0;
while (i < 10) {
body_pos();
i++;}}
\end{lstlisting}
Simultaneously, the loop body is stored as an import object component, to be attached to the WASM module as a function during instantiation.
The original JS loop is replaced by a single binding call that invokes the corresponding exported WASM function. For example:
\begin{lstlisting}
instance.exports.for_pos();
\end{lstlisting}

\noindent\textbf{Rule 9 - Replace Function Definitions (LLM-based):} This rules targets standalone JS function declarations and expressions, including \texttt{FunctionDeclaration}, \texttt{FunctionExpression}, and \texttt{ArrowFunctionExpression} nodes; and replaces them with semantically equivalent AssemblyScript exports. It excludes class methods and focuses on converting top-level or inline-defined functions. 
We exclude class methods because preserving \texttt{this}-dependent semantics and private state across the JS--WASM boundary would require substantial object-model handling (identity, closures, memory layout), beyond the scope of this work.

JS function bodies are highly diverse (nested scopes, closures, higher-order patterns, dynamic typing), making deterministic translation brittle. LLMs generalize across these forms and generate well-typed AssemblyScript. 
 While better results may be achieved with more recent models, our goal was to demonstrate that LLMs can be used for this translation, not to optimize for maximum performance. 

Unlike other rules that rely on structural transformation, this rule uses an LLM to translate a JS function into valid AssemblyScript. We use \verb|Qwen2.5-Coder-14B-Instruct| (Qwen2.5-Coder series)~\cite{hui2024qwen2} with a prompt of the form: \textit{translate the given JS function to AssemblyScript, name it \texttt{func\_def\_123}, export it, and output code only}.
This LLM-generated AssemblyScript code is then compiled and validated before inclusion in the main script. A typical output for a named function looks like:
\begin{lstlisting}
export function add(a: i32, b: i32): i32 {
return a + b;}
\end{lstlisting}

Depending on whether the original was a declaration or expression, the appropriate JS binding code is generated. For named declarations, this is straightforward:
\begin{lstlisting}
let add = instance.exports.add
\end{lstlisting}
For anonymous expressions or arrow functions, a generic reference is used:
\begin{lstlisting}
instance.exports.func_def_123
\end{lstlisting}

Each transformed function is recorded in a dictionary that contains its position and binding code to enable patching.

\noindent\textbf{Rule 10 - Obfuscate Fingerprinting-Relevant MemberExpressions:} This rule targets property accesses (e.g., \texttt{canvas.toDataURL},  \texttt{screen.availHeight}, etc.) including fingerprinting-relevant APIs and replaces them with dynamically reconstructed property names using WASM-exported string fragments. The transformation attempts to obscure key API names by splitting them into two halves and reconstructing them at runtime, making static analysis and detection more difficult.

The property name is split into two (e.g., ``fillText" becomes ``fill" and ``Text").

A pair of AssemblyScript string exports is generated to hold each half:
\begin{lstlisting}
export const f_h_123: string = "fill";
export const s_h_123: string = "Text";
\end{lstlisting}

These are stored in memory to be dynamically reconstructed in JS using a \texttt{getString} function and concatenated:
\begin{lstlisting}
[getString(instance.exports.f_h_123) + getString(instance.exports.s_h_123)]
\end{lstlisting}

Depending on how the property is originally accessed, dot notation (e.g., \texttt{ctx.fillText}) or string literal (e.g., \texttt{ctx["fillText"]}), the replacement logic adapts accordingly. Dot accesses include the leading dot in the replacement range, while bracket-style string keys are reconstructed directly. This is particularly useful where visibility of key API names (such as \texttt{toDataURL}, \texttt{getContext}, or \texttt{platform}) must be hidden or scrambled to reduce traceability by heuristic or signature-based detectors.

\subsection{Datasets}
\label{subsec:datasets}

\noindent\textbf{Real-World JS Corpus.} 
To ensure our evaluation reflects the contemporary deployment practices, we collected a new dataset instead of using an existing datasets (such as DeepFPD \cite{deepfpd}) which may not represent the current state of the web. Therefore, we constructed a large-scale real-world dataset by crawling the May 2025 snapshot of Google’s CrUX Top-1M list\cite{crux}, giving us an up-to-date snapshot of the current landscape of fingerprinting code and enabling comprehensive evaluation in our experiments. 
We implemented the crawler in Python, issuing HTTP requests using the \texttt{requests} library and parsing responses with \texttt{BeautifulSoup}\cite{beatifulsoupuse}. To ensure realism and minimize detection or cloaking, each request included a realistic desktop \texttt{User-Agent} string, standard language and compression headers, and a five-second timeout.
Crawling was distributed across a 64-core node with 8000 parallel workers and a two-second per-host delay. We avoided authentication, form submissions, and JS execution to prevent self-fingerprinting. For each reachable site, we extracted inline scripts and \texttt{<script src>} resources; we additionally visited same-eTLD+1 subdomains discovered in anchor links while excluding third-party domains.

Each script was normalized (UTF-8, LF), deduplicated (SHA-256), and stored in chunked JSON with metadata (origin, capture time, MIME type, status, size, initiator, WASM flag). We then applied the heuristic fingerprinting detection method proposed by Iqbal et al.~\cite{fp_inspector_paper} to label scripts as fingerprinting/non-fingerprinting and classify detected fingerprinting scripts by technique. 
Although fingerprinting payloads evolve rapidly, the underlying structural API properties targeted by these heuristics remain highly consistent, providing a reliable classification foundation for our corpus. 
Our real-world JS corpus consists of \textbf{7,578,653} script files. From this corpus, we heuristically identified \textbf{10,742} scripts containing fingerprinting behavior. Among them, we found a total of 8730, 1780, 1986, and 109 scripts of categories AudioContext, Canvas, WebRTC and Canvas-font, respectively. Because the heuristics from Iqbal et al.\cite{fp_inspector_paper} prioritize high-confidence detection of established techniques, they inherently underestimate the total prevalence of fingerprinting in the wild. Consequently, our real-world corpus overlaps primarily with the Graphics (Canvas), Media (Audio), and Networking (WebRTC) subsets of our evaluation. To address this coverage gap, we utilize the controlled dataset (Table \ref{tab:controlled_dataset}) to supply a broader, representative cross-section of fingerprinting categories. This dual-dataset approach ensures our WASM obfuscation pipeline and the subsequent detection evaluations are tested against both high-volume commercial scripts found in the wild and a wider variety of specialized vectors (e.g., WebGPU, Battery, and WASM-native timing) not targeted by our crawler.

\noindent\textbf{Controlled Dataset.}
While real-world scripts are essential for realism, their obfuscation and nested control flow make post-conversion functional verification difficult to scale. We therefore also use a controlled dataset of paired JS and WASM scripts crafted from known fingerprinting vectors. These scripts display a SHA-256-based fingerprint, enabling direct before/after output comparisons between the original JS and each WASM-converted variant.

The controlled dataset contains 32 JS scripts of 7 different types of fingerprinting techniques. 17 scripts generate stable and unique fingerprint hashes, making them useful for monitoring their behavior over time. The remaining scripts, although they do not produce consistent hashes, still extract valuable information that can assist in device categorization or serve as supplementary features in broader fingerprinting techniques. 
An example of a fingerprinting technique that produces unstable hashes but is still useful is the method proposed by Guri et al.~\cite{guri2025}, which we implement in the WASM Timing Tests category of our controlled dataset. Although this technique does not generate a stable or unique hash, the authors claim that the datapoints it collects can still be used to train a ML classifier that effectively distinguishes between Chromium-based and non-Chromium browsers.

The dataset contains up to four WASM-obfuscated variants per script: one generated by our pipeline, plus three additional variants enabled by the scripts' intentionally simple, fully controlled structure (which broadens testing coverage beyond what a single transformation strategy exercises):
\begin{enumerate}
    \item Pipeline conversion (\cref{sec:cnv_ppl}).

    \item Property-access masking by rewriting dot/bracket access to computed access with split strings stored in WASM (e.g., \lstinline|fillText| $\rightarrow$ \lstinline|"fill"+"Text"|).
    
    \item Whole-script embedding: store the full JS program as a WASM string and execute it via an indirect, WASM-obfuscated \texttt{eval} at runtime.

    \item WASM hashing: replace the JS hashing routine ( \textit{crypto}) with \textit{hash-wasm}~\cite{hashwasm}.

\end{enumerate}

A brief overview of the scripts in the controlled dataset is presented in \cref{tab:controlled_dataset}.

\begin{table}[htb]
\vspace{-0.5cm}
  \scriptsize
  \centering
\caption{Controlled dataset description.}
  \label{tab:controlled_dataset}
  \begin{tabularx}{\linewidth}{@{}lX@{}}
    \toprule
    \textbf{Category} & \textbf{Fingerprinting Methods (variants)} \\
    \midrule
    Environment \& System      & Battery, Datetime Format, Navigator \& Screen, Permissions, Pressure (4 each) \\
    Graphics Rendering         & Canvas, Paint, WebGL, WebGPU (4 each); SVG (3) \\
    Input \& Interaction       & Gamepad, Keyboard, Pointer (4 each) \\
    Media \& Sensors           & Audio (5); MediaCapabilities, MediaDevices, Motion, Sensors (4 each) \\
    Networking                 & Fetch, Network, Timing (4 each); WebRTC (5) \\
    Storage \& Execution       & Filesystem, IndexedDB, ServiceWorker, SharedArrayBuffer, Storage (4 each) \\
    Timing \& Performance      & Performance, UserTiming (4 each); Scheduler (5) \\
    WASM Timing Tests          & Three tests (1); Full test suite (1) \\
    \midrule
    \textbf{Total Scripts} & \textbf{124} \\
    \bottomrule
  \end{tabularx}
\vspace{-0.5cm}
\end{table}

\section{Experimental Setup}
\label{sec:exp_setup}

Our evaluation comprises four experiments aligned with our research questions (\cref{sec:rqs}). We use both our real-world corpus and controlled dataset to assess whether JS-to-WASM obfuscation is feasible at scale and how it affects fingerprinting detectors and defenses.

\noindent\textbf{Dataset Sampling Strategy.} Due to the scale of our real-world corpus, we construct 10 balanced evaluation subsets using a two-stage strategy. Following \cref{subsec:datasets}, we split scripts into fingerprinting and non-fingerprinting pools and form subsets of 500 scripts each (400 fingerprinting, 100 non-fingerprinting). Fingerprinting scripts are selected via stratified sampling across four categories (canvas, WebRTC, canvas-font, AudioContext), yielding 414, 537, 47, and 190 scripts, respectively. Non-fingerprinting scripts are selected via reservoir sampling~\cite{Vitter1985RandomSW}.

\subsection{Experiment 1: Conversion Pipeline Evaluation}
\label{sec:exp-setup-exp1}
This initial experiment quantifies the operational characteristics of our conversion pipeline to establish the viability of transforming JS into WASM-obfuscated code at scale. We measure performance across four dimensions:

\noindent\textbf{Conversion Coverage.}
We report the fraction of original JS characters transformed into WASM-backed bindings: $\text{Coverage}=\frac{\text{Chars}_{\text{transformed}}}{\text{Chars}_{\text{original}}}\times100$.

\noindent\textbf{Conversion Success Rate.}
A script is considered successfully converted if: (i) its generated AssemblyScript compiles with \texttt{asc}; (ii) the modified JS parses into an AST with \texttt{python-esprima}; and (iii) the modified JS executes without runtime exceptions when injected into a test website in headless Chromium. We report $\text{Success Rate}=\frac{N_{\text{pass}}}{N_{\text{total}}}\times100$.

\textbf{Execution Viability vs. Functional Correctness.} We apply two distinct verification criteria based on the dataset. For the real-world corpus, the per-category success rate measures execution viability—verifying only that the converted script compiles, instantiates, and executes without raising a runtime exception. True functional correctness (output-level equivalence) is established exclusively on the controlled dataset: every stable-hash script produced a byte-identical SHA-256 fingerprint to its JS original across all generated WASM variants, and the non-deterministic scripts (e.g., the Guri et al.~\cite{guri2025} WASM Timing Tests) were verified to continue extracting the same measurement set.

\noindent\textbf{Conversion and Validation Time.}
We measure average per-script conversion time (AST traversal, rule application, AssemblyScript generation, WASM compilation) and validation time (parse+execute).

\noindent\textbf{Change in Code Size.}
We compare the size of the original JS file to the total size of the WASM-converted output, which includes (i) the transformed JS, (ii) the compiled .wasm binary, and (iii) any helper glue code. The size is calculated by writing the scripts to .js files and calculating the number of bytes occupied on the disk.
We report both absolute and relative differences to highlight any increase in payload size:
$\text{Relative Increase}=\frac{\text{Size}_{\text{wasm+js}}-\text{Size}_{\text{original\_js}}}{\text{Size}_{\text{original\_js}}}\times100$.

\subsection{Experiment 2: Detection Performance Analysis}
\label{sec:exp-setup-exp2}
This experiment examines whether SOTA fingerprinting detectors in the research literature can identify fingerprinting behavior after our WASM conversion.

\noindent\textbf{Detection Tool Implementation}
We evaluate two representative detectors from the literature, chosen to span the static and hybrid design space:

\noindent\textbf{Tool-1: DeepFPD (Static Detector)}
The original DeepFPD paper~\cite{deepfpd} proposes a deep learning-based fingerprinting detector that leverages features derived from the Abstract Syntax Tree (AST), token sequences, and Control Flow Graphs (CFGs). These features are used to train Long Short-Term Memory (LSTM) and Gated Graph Neural Network (GGNN) models.
We found the provided deep-learning pipeline difficult to reproduce due to deprecated dependencies; the paper also reports that a \textbf{\textit{Random Forest}} trained on the extracted features can outperform the deep models in precision. Motivated by this, we use DeepFPD's feature extraction to build a Random Forest-based detector: tokenized AST elements (e.g., \verb|MemberExpression:screen|) are encoded as a sparse Bag-of-Words matrix via \verb|CountVectorizer| (vocabulary capped at 5000 features).
\smallskip

\noindent\textbf{Tool-2: FP-Inspector (Hybrid Detector)}
FP-Inspector~\cite{fp_inspector_paper} combines static and dynamic features collected via an OpenWPM-based crawler~\cite{englehardt2016online}. Following the authors’ guidelines, we embedded original and WASM-obfuscated scripts into simple local pages (served via a Python HTTP server), crawled them headlessly, and processed the resulting logs using FP-Inspector’s feature extraction pipeline.

We note that DeepFPD and FP-Inspector are among the most widely cited representatives of their respective classes. Crucially, the vulnerability we expose is not specific to these two tools but to their \emph{feature family}: DeepFPD relies on AST-derived token tuples (e.g., \texttt{MemberExpression:screen}); FP-Inspector's static component uses keyword and API-call counts extracted from JS source. FP-Radar~\cite{fp_radar} similarly builds features from temporal sequences of Web-API calls resolved through JS source analysis, and FProbe~\cite{fprobe} uses control- and data-flow features extracted from JS ASTs. All four detectors share a common assumption: that the code under analysis is written in JS and that syntactic or lexical signals are present in the source. WASM obfuscation violates this assumption uniformly, so the evasion patterns we observe on DeepFPD are structurally applicable to any detector whose features are derived from JS source tokens, AST structure, or JS-level call graphs.

\subsection{Experiment 3: Evaluation Against Browser-Based Defenses}

This experiment evaluates whether our WASM-converted scripts can evade defenses integrated into modern browsers and popular anti-fingerprinting browser extensions.

\noindent\textbf{Defense Tools.} We evaluated 10 popular anti-fingerprinting extensions and 5 mainstream browsers (latest stable releases as of July 2025). The full list of extensions and browsers is reported in \cref{tab:browser-eval} and \cref{ext-eval}.

\noindent\textbf{Methodology.} Using our controlled dataset to ensure behavioral consistency, we tested each script against a suite of defenses. A defense is considered bypassed if the script is not blocked, its functionality is not disrupted, and its fingerprinting hash output (for stable scripts) does not change.

\subsection{Experiment 4: Ablation Study on Conversion Rules}
To isolate the contribution of each of our 13 WASM transformation rules, this experiment re-evaluates detection performance by applying only one rule at a time. This ablation study allows us to understand the specific impact of each rule on the potential for detection evasion.

\section{Results}

\subsection{Experiment 1: Conversion Pipeline Evaluation}
We pre-processed sampled real-world scripts to ensure tool compatibility, excluding scripts that \texttt{pyesprima} could not parse and scripts larger than 100~kB (which frequently crashed AST construction). Across all 10 sample subsets, these filters excluded fewer than 3\% of scripts; the exclusion rate did not differ significantly between fingerprinting and non-fingerprinting pools, so it is unlikely to bias the reported success rates.

The performance of our conversion pipeline is summarized in~\cref{tab:cnv_ppl_anls_by_sample} (per-sample) and ~\cref{tab:cnv_ppl_anls_by_cat} (per-category). The pipeline achieved a mean success rate of \textbf{85.76\%} across all samples, with a mean conversion coverage of \textbf{25.01\%}. The 95\% confidence intervals for all metrics are narrow, as shown in~\cref{fig:ci_plot}, indicating low variability and precise estimates across samples. We observed a mean conversion time of \textbf{3.45s} and mean size change of \textbf{24.54\%}.

\begin{figure}[htb]
    \vspace{-0.5cm}
    \centering
    \includegraphics[width=0.90\linewidth]{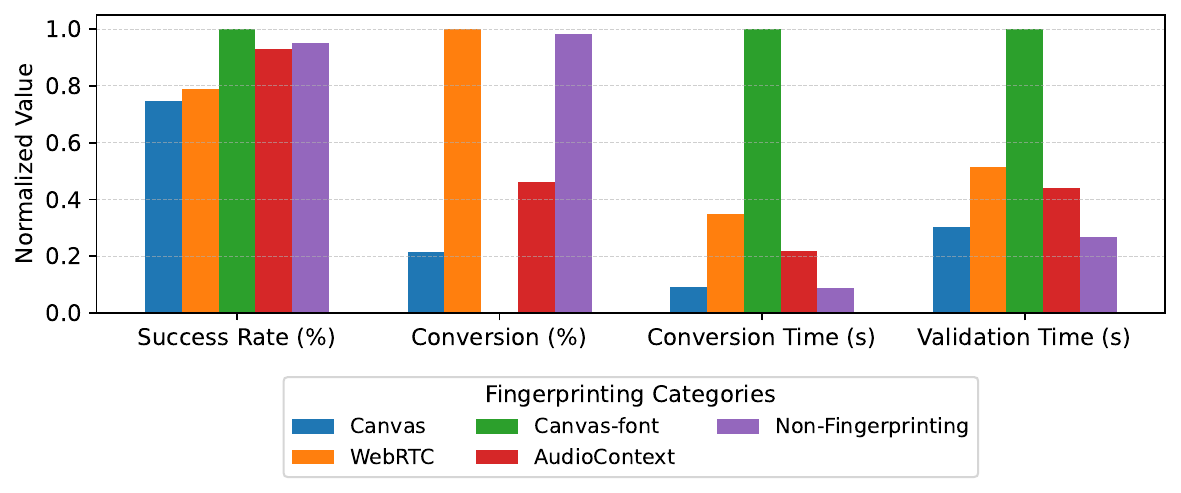}
    \vspace{-0.3cm}
    \caption{Conversion performance by fingerprinting category (normalized).}
    \label{fig:cat_conv_comp}
    \vspace{-0.5cm}
\end{figure}

The conversion success rate varied across fingerprinting categories. A comparison of conversion metrics for categories is provided in \cref{fig:cat_conv_comp}. Canvas-font scripts achieved a perfect 100\%, while Canvas scripts were the most difficult to convert, with a success rate of 74.59\%.
Conversion coverage also differed by category, ranging from a high of 40.31\% for WebRTC scripts to just 0.01\% for Canvas-font. Additionally, Canvas-font scripts had the highest mean conversion time (22.15s), whereas non-fingerprinting scripts were the fastest to convert, with a mean time of 1.99s. We attribute the faster conversion of non-fingerprinting scripts to their simpler structure, whereas fingerprinting scripts often contain extraneous or obfuscated code that increases conversion time.

\begin{table*}[t]
    \centering
    \scriptsize
    \caption{Conversion pipeline performance per sample subset.}
    \label{tab:cnv_ppl_anls_by_sample}
    \begin{tabular}{ccccccc}
    \toprule
      \makecell{\textbf{Sample}\\\textbf{Subset No.}} &
      \makecell{\textbf{Success}\\\textbf{Rate} (\%)} &
      \makecell{\textbf{Conversion}\\\textbf{Coverage} (\%)} &
      \makecell{\textbf{Mean}\\\textbf{Conversion}\\\textbf{Time (s)}} &
      \makecell{\textbf{Mean}\\\textbf{Validation}\\\textbf{Time (s)}} &
      \makecell{\textbf{Size}\\\textbf{Change}} \\
      \midrule
        1 & 84.05\% & 20.21\% & 3.22 & 0.69 & 26.02\% \\
        2 & 86.75\% & 25.43\% & 3.75 & 0.72 & 19.47\% \\
        3 & 88.51\% & 21.46\% & 3.64 & 0.65 & 22.97\% \\
        4 & 84.42\% & 28.40\% & 3.40 & 0.74 & 27.91\% \\
        5 & 86.14\% & 28.01\% & 3.39 & 0.77 & 32.99\% \\
        6 & 85.80\% & 24.45\% & 3.60 & 0.68 & 15.53\% \\
        7 & 85.33\% & 30.01\% & 3.07 & 0.74 & 30.14\% \\
        8 & 85.71\% & 22.10\% & 3.45 & 0.65 & 17.99\% \\
        9 & 85.92\% & 23.91\% & 2.83 & 0.72 & 34.65\% \\
        10 & 84.93\% & 26.07\% & 4.13 & 0.59 & 17.69\% \\\
        \textbf{Mean $\pm$ SD} &
            $85.76\%\pm1.26$ &
            $25.01\%\pm3.20$ &
            $3.45\pm0.36$ &
            $0.70\pm0.05$ &
            $24.54\%\pm6.81$ \\
            \bottomrule
    \end{tabular}
    \vspace{-0.5cm}
\end{table*}

\noindent\textbf{Summary.} Our pipeline converts the large majority of attempted real-world fingerprinting scripts into functional WASM-obfuscated variants (success rate $>85\%$), showing that large-scale JS-to-WASM obfuscation is practical and automatable. 

\subsection{Experiment 2: Detection Performance Analysis}

We evaluate DeepFPD and FP-Inspector on both original and WASM-obfuscated scripts (implementation details in \cref{sec:exp-setup-exp2}).

\noindent\textbf{Tool 1: DeepFPD} 
We evaluate DeepFPD under four scenarios. In Scenarios 1–3, a model trained on DeepFPD’s original dataset (10,240 non-fingerprinting and 343 fingerprinting scripts) is tested on three datasets (summarized in~\cref{tab:deepfpd_1st_3_exp}). WASM obfuscation severely impaired detection on DeepFPD's own test set (Scenario 1), dropping recall from \textbf{77.78\%} to \textbf{44.44\%}. Conversely, the real-world dataset (Scenario 2) experienced only a modest recall drop (86.40\% to 81.28\%). The controlled dataset (Scenario 3) showed near-complete failure.
\begin{table}[htb]
  \vspace{-0.5cm}
  \centering
  \scriptsize
  \caption{DeepFPD performance across three experiments.}
  \label{tab:deepfpd_1st_3_exp}
  \begin{adjustbox}{max width=\columnwidth}
  \begin{tabular}{lcccccc}
    \toprule
    \multirow{2}{*}{\textbf{Scenario}} &
    \multicolumn{2}{c}{\textbf{Accuracy}} &
    \multicolumn{2}{c}{\textbf{Precision}} &
    \multicolumn{2}{c}{\textbf{Recall}} \\
    & JS & WASM & JS & WASM & JS & WASM \\
    \midrule
    1. DeepFPD testset               & 97.18\% & 95.92\% & 73.68\% & 72.73\% & 77.78\% & 44.44\% \\
    2. Real-world dataset    & 91.52\% & 88.54\% & 99.42\% & 99.70\% & 86.40\% & 81.28\% \\
    3. Controlled dataset    &  6.67\% &  1.06\% & 100.00\% & 100.00\% &  6.67\% &  1.06\% \\
    \bottomrule
  \end{tabular}
\end{adjustbox}
\vspace{-0.5cm}
\end{table}

In \emph{Scenario 4} (WASM-augmented retraining), we test whether this performance drop stems from a lack of WASM training data by retraining DeepFPD using WASM-augmented scripts. Adding only category-specific fingerprinting scripts destabilized the classifier: Canvas/WebRTC augmentation plummeted WASM accuracy to $\approx 6\%$ ($\approx 5.7\%$ precision, 100\% recall), while AudioContext augmentation reduced both JS and WASM accuracy and precision to 5.64\% (100\% recall). However, a balanced augmentation (414 Canvas, 537 AudioContext, 190 WebRTC fingerprinting, and 525 non-fingerprinting scripts) restored stability, achieving 98.12\% accuracy, 87.50\% precision, and 77.78\% recall across both JS and WASM test sets. Using \texttt{treeinterpreter}~\cite{treeinterpreter}, we computed per-feature contributions to the fingerprinting class (\cref{tab:deefpd_ftr_impt}). The model relies heavily on high-signal AST tuples tied to fingerprinting APIs.

\begin{table}[htb]
    \vspace{-0.5cm}
    \centering
    \scriptsize
    \caption{DeepFPD model feature importance.}
    \label{tab:deefpd_ftr_impt}
    \begin{tabular}{lc}
        \toprule
        \textbf{AST Tuple} & \textbf{Feature Importance} \\
        \midrule
        MemberExpression:screen        & 0.0214 \\
        MemberExpression:fillText      & 0.0192 \\
        CallExpression:canvas          & 0.0186 \\
        MemberExpression:language      & 0.0145 \\
        MemberExpression:localStorage  & 0.0118 \\
        MemberExpression:appName       & 0.0116 \\
        MemberExpression:platform      & 0.0110 \\
        MemberExpression:fillStyle     & 0.0106 \\
        MemberExpression:colorDepth    & 0.0105 \\
        MemberExpression:fillRect      & 0.0089 \\
        \bottomrule
    \end{tabular}
    \vspace{-0.5cm}
\end{table}

\begin{figure}[htb]
    \vspace{-0.5cm}
    \centering
    \includegraphics[width=0.70\linewidth]{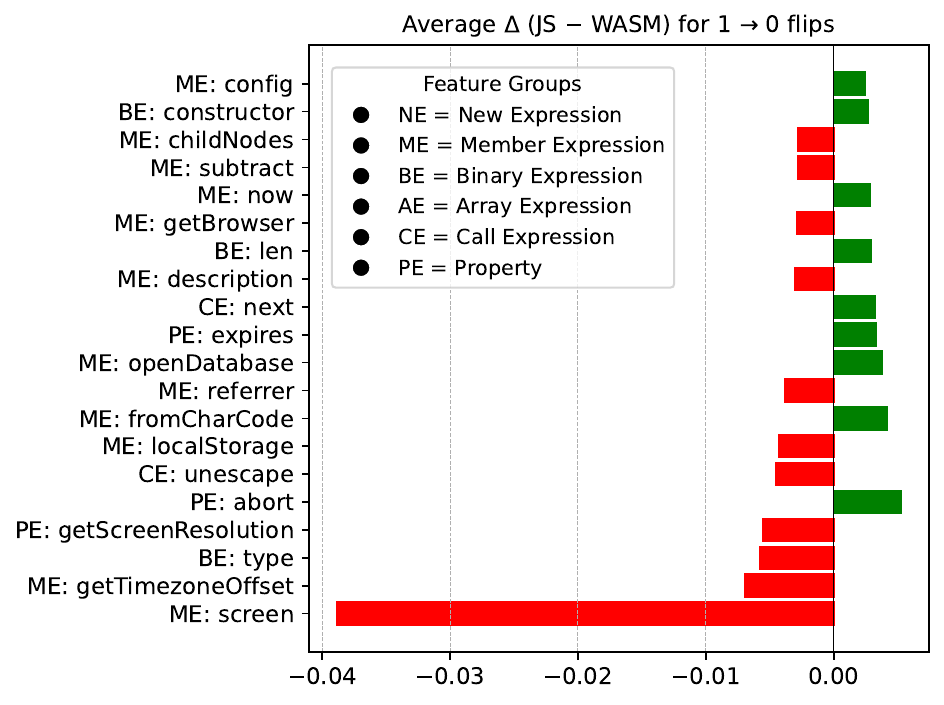}
    \vspace{-0.3cm}
    \caption{Aggregated delta plot of feature contribution changes.}
    \label{fig:aggregrated_delta_deepfpd_testset}
    \vspace{-0.5cm}
\end{figure}

Delta plots (\cref{fig:4_delta_deepfpd_testset,fig:aggregrated_delta_deepfpd_testset}) reveal that obfuscation sharply reduces the predictive weight of features like  \texttt{MemberExpression:getTimezoneOffset}, \texttt{MemberExpression:screen}, \texttt{BinaryExpression:type}, and \newline\texttt{Property:getScreenResolution}. This explains the disproportionate recall drop on the DeepFPD test set, which heavily relies on these easily suppressed features. In contrast, our real-world dataset retained classification accuracy because it relies more on canvas rendering and plugin properties (\texttt{canvas}, \texttt{font}, \texttt{fillRect}, \texttt{plugins}) that remain relatively unaffected by WASM obfuscation. Ultimately, while our \textit{greedy} black-box obfuscation strategy yielded mixed results, an adversary with internal detector knowledge could perform \textit{targeted} conversions on these high-importance features to maximize evasion.

\noindent\textbf{Tool 2: FP-Inspector}
The hybrid static-dynamic detector FP-Inspector failed to process WASM-converted scripts. Its crawler did not capture or include these scripts in the extracted features, preventing the detection pipeline from operating as intended.
This behavior stems from a critical toolchain dependency: FP-Inspector relies on Firefox 52 ESR, a version in which WASM support is disabled. As a result, the detector is not merely evaded but rendered ineffective in the presence of WASM-based implementations. This exposes a broader fragility, where the viability of a detection system depends on compatibility with evolving browser technologies.
While FP-Inspector continues to function on legacy JS-based fingerprinting scripts, it becomes unusable when fingerprinting logic is expressed through WASM. This underscores the importance of maintaining compatibility with modern web standards in dynamic fingerprinting defenses.
To isolate behavior under a WASM-compatible setting, we next build a modern dynamic monitor.

\noindent\textbf{Tool 3: Dynamic Analysis with Modern Software Stack}
To evaluate the robustness of dynamic fingerprinting detection under WASM-based obfuscation, we implement our own API-level monitoring tool using \texttt{Playwright} with the latest Chrome browser. The tool logs all invocations of fingerprinting relevant APIs; covering both method calls and property accesses matching the set of APIs instrumented by FP-Inspector. We also re-implement FP-Inspector's feature-extraction logic to ensure feature compatibility.
For each script, we generate a HTML page containing the required DOM structure (with html tags containing all ids and classes required by the script), host it locally via a lightweight Python HTTP server, and collect API-execution traces using our tool. We extract dynamic features from both the original JS and the WASM-obfuscated versions.

\begin{figure}[htb]
    \vspace{-0.5cm}
    \centering
    \includegraphics[width=0.85\linewidth]{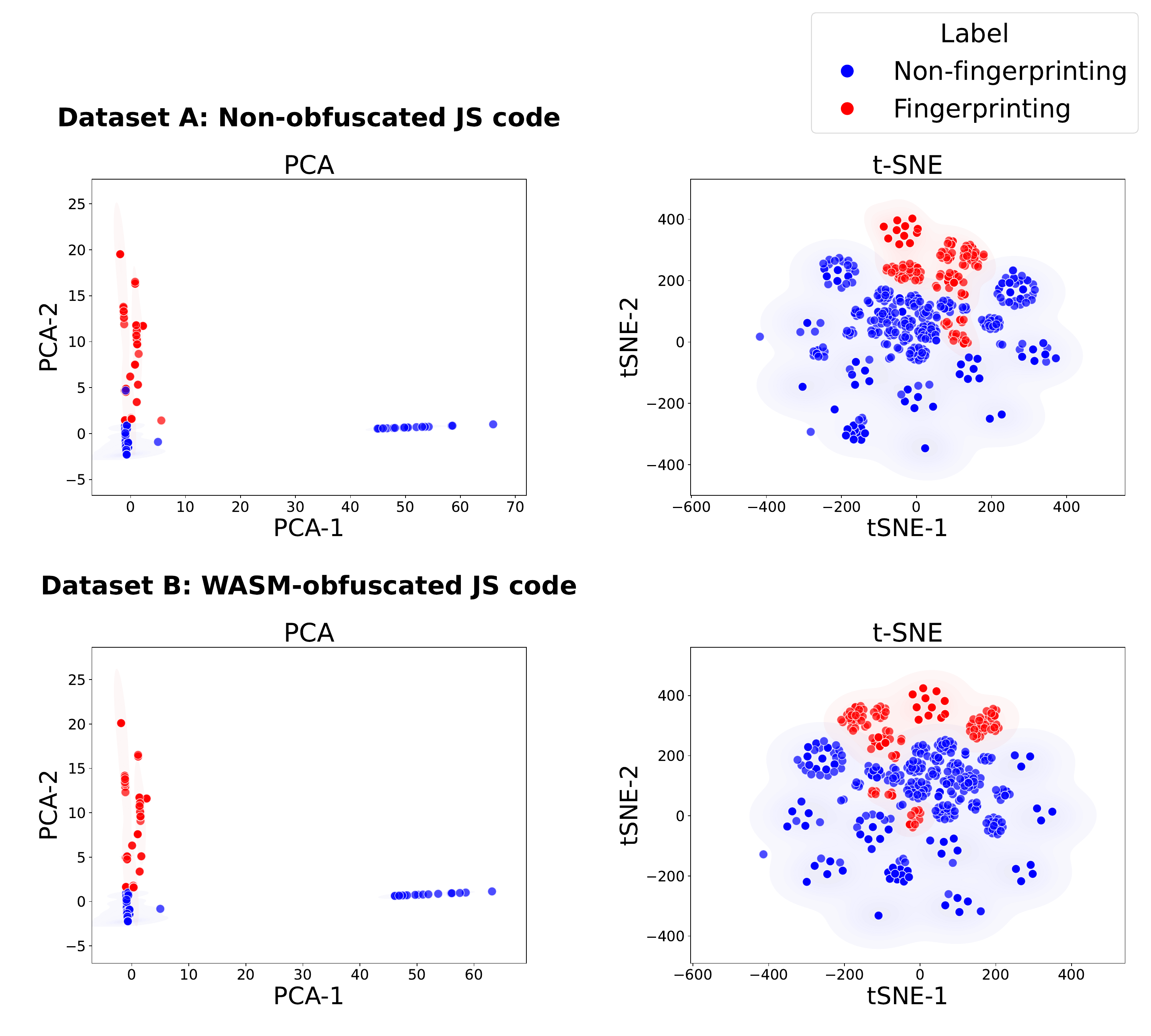}
    \vspace{-0.3cm}
    \caption{PCA and t-SNE of dynamic API-call features for non-obfuscated (A) and WASM-obfuscated (B) scripts.}
    \label{fig:dataset_compare}
    \vspace{-0.5cm}
\end{figure}

We then train six ML classifiers (\texttt{Random Forest}, \texttt{Logistic Regression}, \texttt{Linear SVM}, \texttt{Gradient Boosting}, \texttt{KNN}, \texttt{Naive Bayes}) on non-obfuscated JS traces and test on WASM-obfuscated traces, and vice versa. In all cases, the classifiers maintain near-perfect accuracy (see \cref{jsvwasm}), indicating that dynamic API-level behavior remains highly distinctive even after WASM conversion. The feature-space visualizations in \cref{fig:dataset_compare} further show that although WASM obfuscation induces some structural differences, the fingerprints remain well separable. These results demonstrate that monitoring runtime API calls, along with their argument and return-value patterns, offers a robust and resilient way to detect fingerprinting behavior. This stands in contrast to static code-analysis approaches, which are weakened by WASM-based obfuscation.

\noindent\textbf{Summary.} Static detectors can be degraded by WASM-based obfuscation, and tooling choices can make hybrid detectors fragile (e.g., legacy stacks without WASM support). In contrast, dynamic, API-level monitoring with a modern software stack remains robust for detecting WASM-obfuscated fingerprinting.

\subsection{Experiment 3: Evaluation of Anti-Fingerprinting Extensions and in-built Browser Defenses}

In contrast to the SOTA detectors in the research literature, we found that popular commercial in-browser defenses and browser extensions were largely unaffected by our WASM obfuscation technique. As shown in~\cref{ext-eval} and~\cref{tab:browser-eval}, all 10 tested browser extensions (including CanvasBlocker, Privacy Badger, and uBlock Origin) and all 5 browser configurations (including Firefox with Enhanced Tracking Protection and Brave with fingerprinting protection) remained unaffected by WASM obfuscation.

Our analysis reveals that this resilience arises from a fundamentally different defense model. Tools such as \textbf{CanvasBlocker} and browsers like \textbf{Firefox} (in Strict Mode) do not rely on static or dynamic code analysis to detect fingerprinting. Instead, they use API-level defenses by intercepting and spoofing sensitive outputs (e.g., from canvas or audio APIs), or by blocking access to specific browser interfaces. These defenses are agnostic to the origin or structure of the invoking code and, therefore, remain effective regardless of whether the fingerprinting logic is implemented in JS or WASM.
Similarly, extensions such as \textbf{Disconnect}, \textbf{Privacy Badger}, and \textbf{uBlock Origin} rely on domain-level blocking to prevent tracking. Since our evaluation was conducted on locally hosted pages, these tools were not triggered and did not intervene. This reflects a limitation in their operational scope rather than a weakness against WASM obfuscation.

\noindent\textbf{Summary.} Anti-fingerprinting extensions and built-in browser defenses were unaffected by our JS-to-WASM obfuscation, consistent with API-level defenses being agnostic to code structure. However, WASM-native techniques that avoid monitored APIs (e.g., timing-based micro-benchmarks) may still bypass API-centric defenses.

\subsection{Experiment 4: Ablation Study on Conversion
Rules}
The ablation study, summarized in \cref{tab:ablation_sumamry}, assessed the individual contribution of each of our 13 conversion rules. When evaluating the transformed scripts using DeepFPD, we observed that no single rule alone was sufficient to produce a notable evasion effect. In fact, applying only one rule resulted in no change in classification metrics between the JS or WASM versions of the DeepFPD test set.  
During conversion, \texttt{replace\_with\_regex} achieved the highest stand-alone success rate (\(54.10\%\)) while contributing a modest \(3.25\%\) size increase and \(0.0004\%\) coverage. In contrast, \texttt{replace\_canvas\_api\_calls} increased size by nearly \(200\%\) but yielded a lower success rate (\(16.79\%\)). \texttt{replace\_func\_defs} performed worst overall (3.40\,s per script; 9.70\% success), largely due to LLM-based translation overhead.

\begin{table*}[htb]
\vspace{-0.5cm}
\centering
\scriptsize        %
\caption{Summary of ablation study on conversion rules.}
\label{tab:ablation_sumamry}
\begin{tabularx}{\textwidth}{L c c c c c}
\toprule
\textbf{Conversion Rule} &
\makecell{\textbf{Success}\\\textbf{Rate (\%)}} &
\makecell{\textbf{Conversion}\\\textbf{Coverage (\%)}} &
\makecell{\textbf{Mean}\\\textbf{Conversion}\\\textbf{Time (s)}} &
\makecell{\textbf{Mean}\\\textbf{Validation}\\\textbf{Time (s)}} &
\makecell{\textbf{Size}\\\textbf{Difference (\%)}} \\
\midrule
obfuscate\_functions                       &  33.21 & 0        & 1.3774 & 0.5546 & 3.01 \\
replace\_callee                            &  41.79 & 0        & 1.0015 & 0.5389 & 2.96 \\
replace\_canvas\_api\_calls                &  16.79 & 0.1255   & 0.6949 & 0.6314 & 199.84 \\
replace\_class\_defs                       &  52.99 & 0.0002   & 0.7069 & 0.6010 & 2.54 \\
replace\_float\_arrays                     &  23.88 & 0.0005   & 0.6744 & 0.5027 & 25.77 \\
replace\_for\_loops                        &  24.63 & 0        & 0.8179 & 0.3823 & 11.27 \\
replace\_function\_calls\_\allowbreak with\_no\_return
 &  24.25 & 0.3335   & 0.6290 & 0.5334 & -45.62 \\
replace\_func\_defs                        &   9.70 & 0.2154   & 3.4002 & 0.3589 & 21.48 \\
replace\_if\_else                          &  19.78 & 0.0147   & 0.7282 & 0.3824 & 11.30 \\
replace\_int\_arrays                       &  23.13 & 0.0007   & 0.6694 & 0.4910 & 25.71 \\
replace\_literals\_recursive               &  21.64 & 0.0216   & 0.6909 & 0.4844 & 13.26 \\
replace\_obf\_screen                       &  53.73 & 0.0003   & 1.0849 & 0.5796 & 3.30 \\
replace\_while\_loops                      &  51.49 & 0.0021   & 1.7270 & 0.5618 & 2.95 \\
replace\_with\_regex                       &  54.10 & 0.0004   & 0.6229 & 0.6175 & 3.25 \\
\bottomrule
\end{tabularx}
\vspace{-0.5cm}
\end{table*}

\noindent\textbf{Summary.} The ablation study indicates that the evasion observed in Experiment 2 is not attributable to any single rule; instead it emerges from combining multiple transformations. This supports the need for a broader, greedy obfuscation strategy that perturbs several script aspects.

\section{Discussion}
\label{sec:discussion}
\vspace{-0.1cm}
Our empirical analysis reveals a stark divide: feature-based academic detectors are vulnerable to WASM obfuscation, whereas practical, in-browser API-level defenses are immune; highlighting a disconnect between academic analysis and deployed defenses. We also find, while dynamic detection failures (e.g., FP-Inspector) stem largely from engineering gaps in outdated toolchains, static analysis of WASM faces intrinsic language hurdles. JavaScript retains rich syntactic structures, variable names, and explicit control-flow graphs in its AST. In contrast, WASM is a stack-based binary format that strips away these high-level semantic cues. Consequently, porting JS-centric static heuristics to WASM requires fundamentally new analysis paradigms capable of inferring browser API intent directly from low-level memory operations.

\noindent\textbf{Limitations.}
Our evaluation covered DeepFPD (static) and FP-Inspector (hybrid); the latter's failure was a toolchain-rot issue rather than a test of its detection logic. While we argue (\cref{sec:exp-setup-exp2}) that the vulnerability generalizes to detectors sharing the same JS-centric feature families, it is also worth acknowledging dynamic taint-tracking systems such as FPTracer\cite{boussaha24fptracer}. We did not directly evaluate these because their expected evasion would likely stem from technical implementation gaps (the inability to propagate taint across the JS-WASM boundary) rather than a fundamental conceptual limitation.

Further, while our \textit{greedy} translation strategy stress-tested defenses by maximizing WASM footprint, it operates as a black-box. A feature-aware approach that selectively obfuscates high-impact features may yield stronger evasion, motivating future exploration of more targeted strategies.
Additionally, our reliance on the structural heuristics from  Iqbal et al.\cite{fp_inspector_paper} to build the real-world corpus means our wild dataset underestimates the true volume and diversity of deployed fingerprinting. Future at-scale measurements must incorporate updated, wider-net heuristics to capture emerging hardware and timing-based tracking vectors.

\noindent\textbf{Future Work.} Our findings also highlight critical avenues for future research in developing more resilient defenses. One key direction is the development of WASM-aware detectors that can operate directly on WebAssembly binaries to identify suspicious instruction patterns, or apply dynamic techniques that trace execution across the JS and WASM boundary. Complementing this, our results suggest that hybrid defense models, which combine structural code analysis with dynamic API-level interception, offer the most realistic path forward. By integrating behavioral cues with code-level features, these models could detect fingerprinting even when advanced obfuscation is applied.

\noindent\textbf{Conclusion}
This paper provides the first large-scale measurement of how JS-to-WASM obfuscation affects deployed browser fingerprinting defenses. We showed that WASM obfuscation measurably weakens academic, feature-based detectors, while API-level defenses remain robust; a divergence that underscores the need for WASM-aware, cross-language detection in the next generation of tools.

\begin{credits}

\subsubsection{\discintname}
The authors have no competing interests to declare that are relevant to the content of this article.
\end{credits}

\section*{Ethical Considerations}
\label{sec:ethics}

We analysed only JavaScript that was publicly available in the May 2025 CrUX Top‑1M snapshot; no personal or user‑generated data were collected, so the study is outside the scope of human‑subjects research and required no IRB review. Throughout the crawl we respected public‑content boundaries, added a two‑second delay between requests, and avoided any authentication or form submission, following standard ethical‑crawling practice.
Our primary contribution, the JS-to-WASM conversion pipeline, did not reveal any new security vulnerabilities in the tested commercial browser extensions or built-in defenses. Consequently, no responsible disclosure was necessary. We note that our evaluation of a pre-existing, publicly documented WASM native fingerprinting technique from prior research did demonstrate an evasion of these tools. As this technique is already in the public domain, it does not fall under the standard protocol for disclosing a newly discovered vulnerability. Its inclusion serves to highlight the limitations of current API-centric defenses for the academic and security communities.

\section*{Open Science} To benefit future research, the code, the real-world JS-corpus, and the controlled dataset has been publicly released\cite{githubGitHubNazanaza2970WASM_cloak_browser_fingerprinting}.
\appendix
\label{appendix}

 \section{Appendix}
\subsection{Remainder of Translation Rules}
\label{sec:remaining_rules}

\noindent\textbf{\textit{Rule 2 - Obfuscate Sensitive Function Calls:}} Replaces direct invocations of sensitive functions (\texttt{eval}, \texttt{Function}, \texttt{atob}, \texttt{unescape}) with WASM equivalents where names are dynamically resolved from linear memory, preventing static signature detection. For example, \verb|eval(arg0)| is transformed into an AssemblyScript export (\verb|export let eval_123 = "eval";|) and reconstructed in JS via \verb|globalObject[getString(ptr)](arg0)|. It explicitly skips object constructors (e.g., \texttt{new ActiveXObject}) to preserve semantics.

\noindent\textbf{\textit{Rule 3 – Replace Integer and Float Arrays:}} Replaces arrays initialized exclusively with numbers by storing their contents in WASM memory. The AST generates AssemblyScript that allocates memory (\verb|__alloc(length * sizeof<i32>())|) and stores the values. JS binding code reconstructs the array at runtime via \verb|new Int32Array(instance.exports.memory.buffer, ptr, length)|. Mixed-type arrays or \verb|new Array()| instantiations are ignored.

\noindent\textbf{\textit{Rule 4 – Replace If-Else Statements:}} Offloads simple \texttt{if-else} constructs lacking disruptive control flow (\texttt{return}, \texttt{throw}, \texttt{break}) to WASM. The condition, \texttt{then}, and \texttt{else} branches are mapped to a dispatch function (e.g., \verb|if (cond == 1) {$imp1();} else {$imp2();}|) with the branches passed as JS imports. JS binding passes a binary flag to evaluate the construct.

\noindent\textbf{\textit{Rule 6 – Replace While Loops:}} Extracts \texttt{while} loop conditions and bodies, moving the termination logic into an AssemblyScript \texttt{if-else} block. Parameters are dynamically derived from variables used in the condition. The condition and body are passed as import object components, and JS invokes the exported loop wrapper (\verb|instance.exports.f_pos();|).

\noindent\textbf{\textit{Rule 7 – Replace Function Calls Without Return Values:}} Targets standalone function calls in expression statements whose return values are unused. The call is wrapped in an AssemblyScript export (e.g., \verb|export function f_pos(): void {impFunc_pos();}|), with the original function body passed as an import. The JS AST is patched to invoke the WASM wrapper.

\noindent\textbf{\textit{Rule 8 – Replace Class Definitions:}} Offloads \texttt{ClassDeclaration} or \texttt{ClassExpression} nodes by storing the entire class definition as a raw WASM string export (e.g., \verb|export let class_pos = "class MyClass {...}";|). Because WASM cannot pass complex objects directly, JS binding logic reads the string via \texttt{getString()} and injects it back into the DOM using a dynamically created \texttt{<script>} element.

\noindent\textbf{\textit{Rule 11 - Obfuscate Dynamic Code Generation Calls:}} Lifts dynamic constructs (e.g., \texttt{canvas()} calls or fingerprinting member expressions like \texttt{screen.availHeight}) into WASM memory as obfuscated string constants. For instance, \texttt{eval(canvas())} is split into exported strings and reconstructed via an indirect call: \verb|window[getString(e_call)][getString(c_str)]|.

\noindent\textbf{\textit{Rule 12 – Regex-Based Obfuscation of Fingerprinting Strings:}} Replaces hard-coded occurrences of the string ``canvas'' (excluding JSON keys) with dynamically reconstructed equivalents. It emits split WASM string exports (e.g., \verb|"can"| and \verb|"vas"|) and generates JS to concatenate them via memory lookups.

\noindent\textbf{\textit{Rule 13 – Obfuscate MemberExpressions with screen Using Split Identifiers:}} Targets obfuscated \texttt{screen} and \texttt{canvas} member expressions often hidden behind hex-encoded strings (e.g., \verb|"\x73\x63\x72\x65\x65\x6E"|). It emits split AssemblyScript string exports (\verb|"scr"| and \verb|"een"|), allowing the JS binding to reconstruct the object name at runtime before property access.

\section{Additional Results}
\label{sec:app-results}

\noindent\textbf{Bypassing Defenses with WASM-Native Techniques:}
While our JS-to-WASM conversion did not evade API-level defenses, we evaluated a fully WASM-native fingerprinting technique~\cite{guri2025} that bypasses them entirely. By performing timing-based micro-benchmarking directly within WASM bytecode, it avoids monitored APIs (e.g., Canvas, WebGL). In our testing, it successfully fingerprinted the browser even with defenses like CanvasBlocker and Firefox Strict Mode enabled, highlighting that current API-centric defenses remain blind to microarchitectural or timing side-channels originating from within WASM.

\subsection{Conversion Metrics}
\begin{figure*}[htb]
    \centering
    \includegraphics[width=0.95\textwidth]{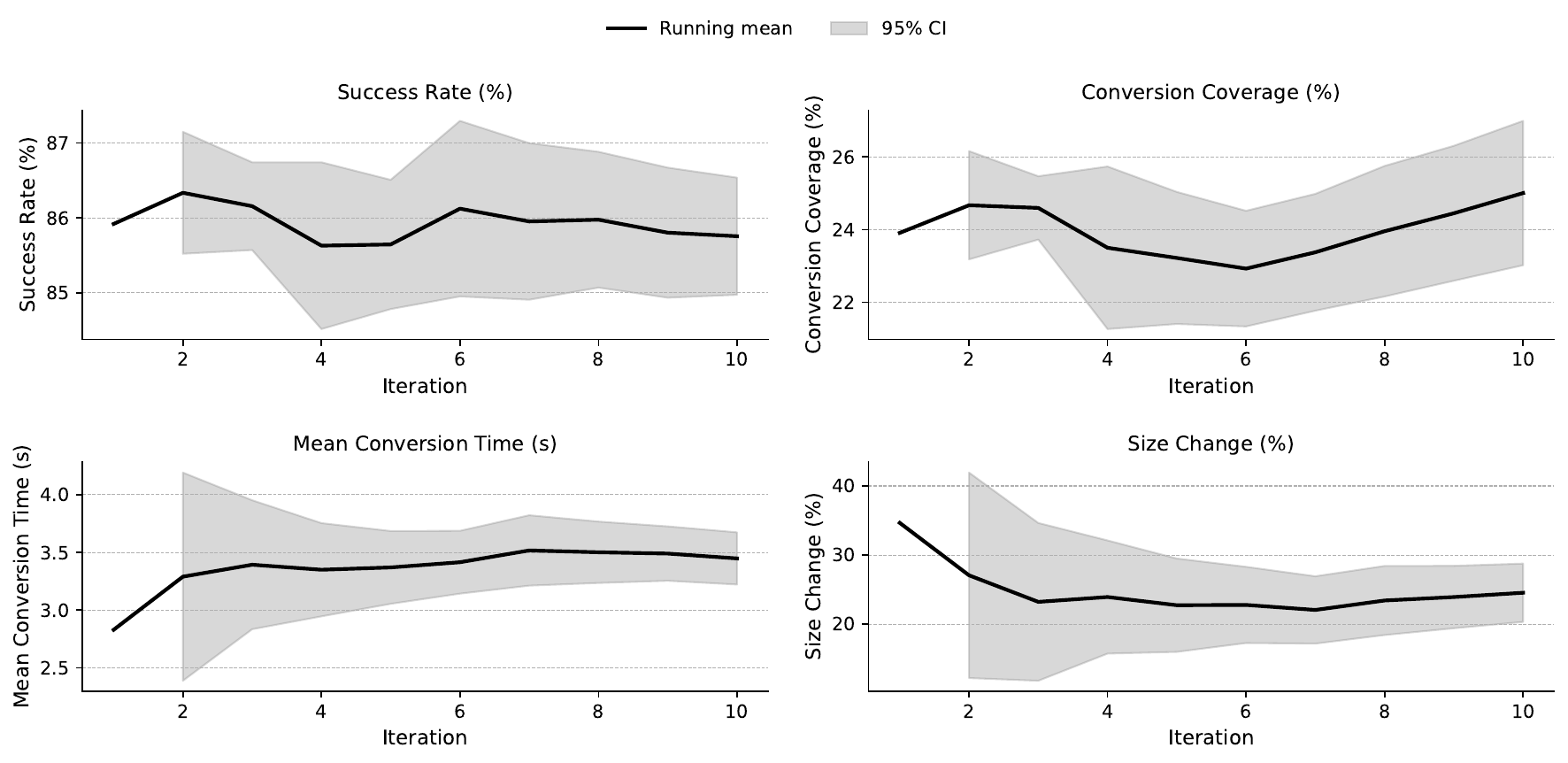}
    \caption{Running mean and 95\% confidence intervals of conversion metrics on the real-world dataset.}
    \label{fig:ci_plot}
\end{figure*}

\Cref{fig:ci_plot} shows the running mean and 95\% confidence intervals for four key metrics: Success Rate, Conversion Coverage, Mean Conversion Time, and Size Change for the real-world dataset.

\subsection{Cross-Environment Classifier Evaluation
\label{jsvwasm}}
To evaluate feature robustness across execution environments, we trained six off-the-shelf classifiers (Random Forest, Logistic Regression, Linear SVM, Gradient Boosting, K-Nearest Neighbors, and Gaussian Naive Bayes) on JS traces and tested them on WASM traces, and vice versa. Performance was highly symmetric: five models achieved 1.0000 for accuracy, weighted precision, and weighted recall in both transfer directions. Gaussian Naive Bayes achieved 0.9900 across all metrics. These bidirectional results demonstrate that dynamic API-level behavior remains robust regardless of the underlying implementation language, rendering tabular reporting redundant.

\subsection{DeepFPD Model Analysis}
\begin{figure}[!ht]
    \centering
    \includegraphics[width=0.8\textwidth]{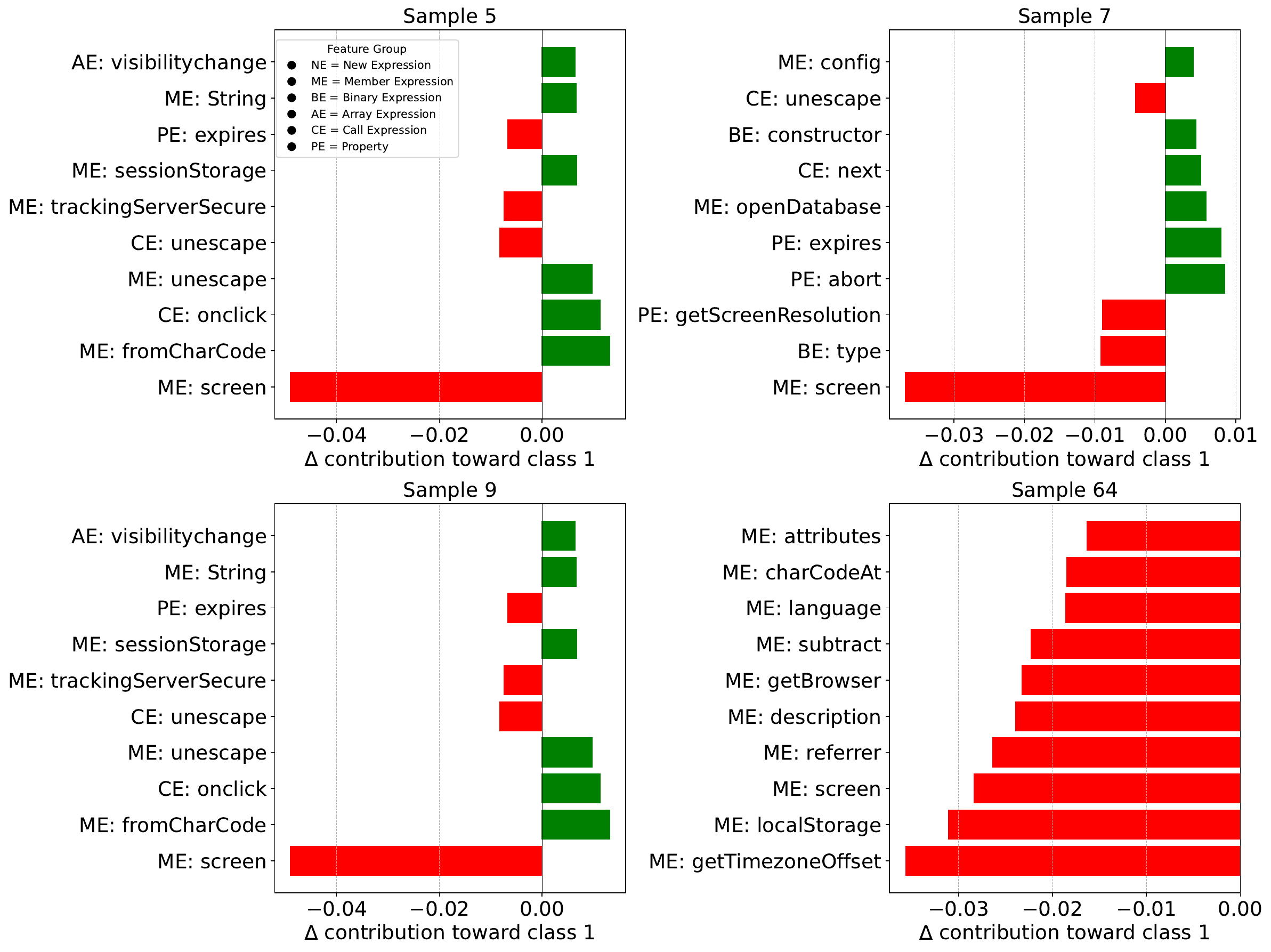}
    \caption{Change in DeepFPD feature contributions for representative test samples after WASM obfuscation.}
    \label{fig:4_delta_deepfpd_testset}
\end{figure}

\begin{figure}[!ht]
    \centering
    \includegraphics[width=0.8\textwidth]{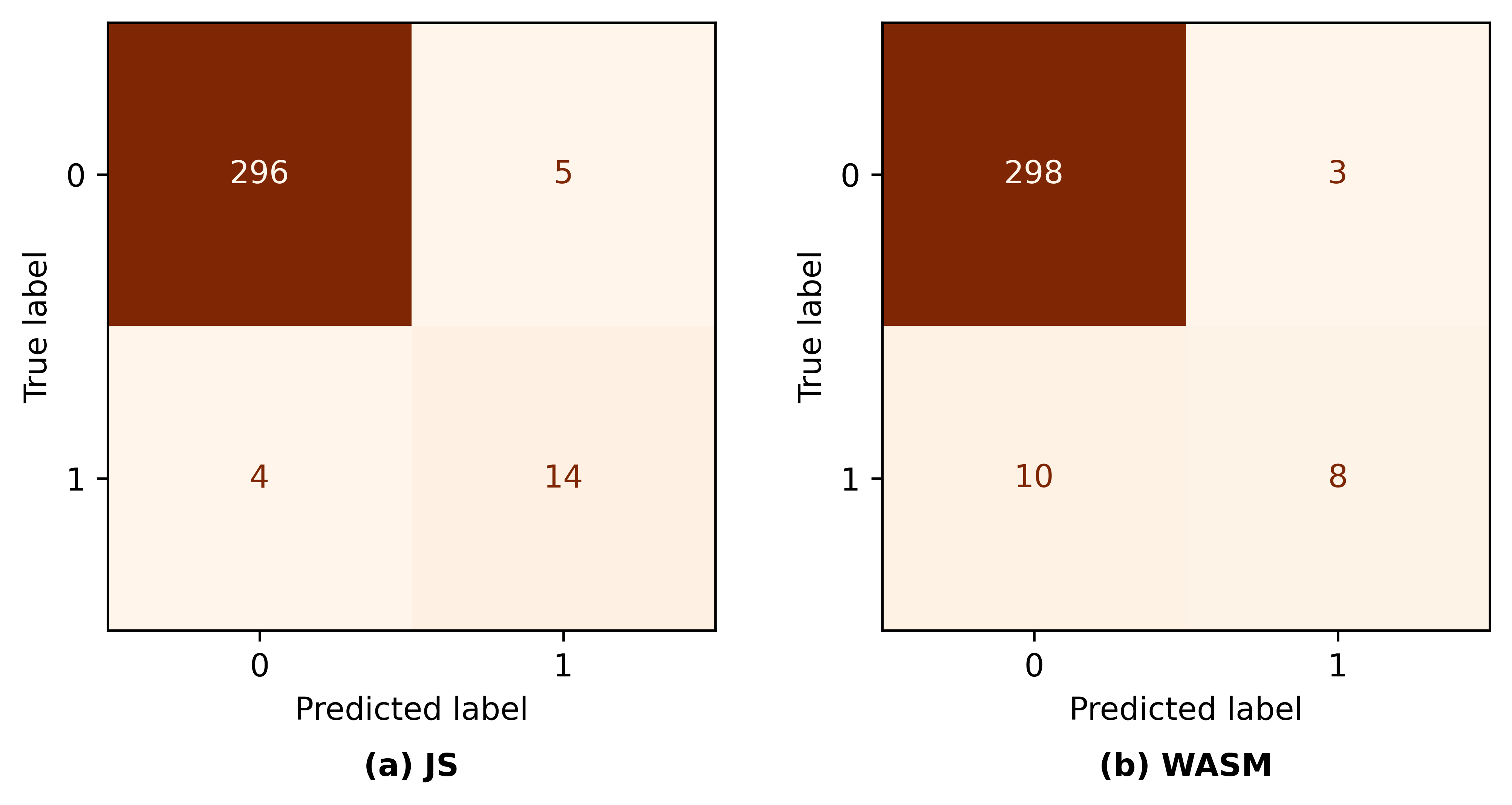}
    \caption{Confusion matrices for DeepFPD predictions on original JS and WASM-obfuscated test sets.}
    \label{fig:deepfpd_conf_mat}
\end{figure}

We visualize the shift in feature attribution scores for representative samples from the DeepFPD test set in \cref{fig:4_delta_deepfpd_testset}. \cref{fig:deepfpd_conf_mat} provides confusion matrices comparing classification outcomes on JS vs. WASM-obfuscated scripts.

\subsection{Conversion Performance by Category}
\Cref{tab:cnv_ppl_anls_by_cat} reports pipeline-level performance across fingerprinting categories.

\begin{table}[htbp!]
    \centering
    \scriptsize
    \caption{Conversion pipeline performance per category.}
    \label{tab:cnv_ppl_anls_by_cat}
    \begin{tabular}{lccccc}
    \toprule
      \textbf{Category} & \textbf{Total} & \makecell{\textbf{Success}\\\textbf{Rate} (\%)} & \makecell{\textbf{Conversion}\\\textbf{Coverage} (\%)} & \makecell{\textbf{Mean}\\\textbf{Conversion}\\\textbf{Time (s)}} & \makecell{\textbf{Mean}\\\textbf{Validation}\\\textbf{Time (s)}} \\
      \midrule
      Canvas Fingerprinting       & 555 & 74.59 &  8.65 &  2.0249 & 0.6438 \\
      WebRTC Fingerprinting       & 242 & 78.93 & 40.31 &  7.6911 & 1.0922 \\
      Canvas-font Fingerprinting  &  47 &100.00 &  0.01 & 22.1537 & 2.1263 \\
      AudioContext Fingerprinting & 579 & 92.92 & 18.66 &  4.8480 & 0.9403 \\
      Non-Fingerprinting          & 553 & 95.30 & 39.59 &  1.9892 & 0.5693 \\
      \midrule
      \textbf{Mean $\pm$ SD}      & --  & 88.35$\pm$10.99 & 21.44$\pm$18.14 & 7.74$\pm$8.39 & 1.07$\pm$0.63 \\
      \bottomrule
    \end{tabular}
\end{table}

\subsection{Summary of Browser-Level Defenses and Extension Testing}
\Cref{tab:browser-eval} covers the five major browsers and their built-in fingerprinting defenses. 

\begin{table}[H]
\centering
\scriptsize
\caption{Summary of browser-level defenses against WASM obfuscation.}
\label{tab:browser-eval}
\setlength{\tabcolsep}{6pt}
\begin{tabular}{@{}llc@{}}
\toprule
\textbf{Browser} & \textbf{Built-in Fingerprinting Defense} & \textbf{Affected by WASM Conversion?} \\
\midrule
Brave\cite{brave2025}     & Partial (Script Blocking) & No \\
Firefox\cite{firefox2025} & Yes (Strict Mode)         & No \\
Chrome\cite{chrome2025}   & No                        & No \\
Edge\cite{edge2025}       & No                        & No \\
Opera\cite{opera2025}     & No                        & No \\
\bottomrule
\end{tabular}
\end{table}

\subsection{Evaluation of Anti-Fingerprinting Extensions \label{ext-eval}}
We evaluated ten popular anti-fingerprinting browser extensions to assess their resilience to WASM obfuscation. None were affected by the conversion, though this resilience stems from differing underlying mechanisms:

\begin{itemize}[leftmargin=*,nosep]
    \item \textbf{API Spoofing and Interception:} Extensions that hook or spoof browser APIs (CanvasBlocker\cite{canvasblocker2025}, Chameleon\cite{chameleon2025}, Canvas Fingerprint Defender, Trace\cite{trace2025}) remained completely effective. Because WASM must bridge to JS to access Web APIs, these interceptions successfully disrupted payloads regardless of the execution layer.
    \item \textbf{Network and Domain Blocking (Out of Scope):} Origin-based or domain blocklisting tools (uBlock Origin\cite{ublockorigin2025}, Privacy Badger\cite{privacybadger2025}, Disconnect\cite{disconnect2025}) were not triggered in our local test environment. WASM obfuscation masks payload semantics but cannot bypass network-level domain filtering.
    \item \textbf{Orthogonal or Incompatible Tools:} Random User-Agent Switcher\cite{randomuaswitcher2025} modifies HTTP headers independent of script execution. JShelter\cite{jshelter2025} and Browser Fingerprint Protector\cite{browserfingerprintprotector2025} failed to parse the scripts or function in our test setup. 
\end{itemize}

\section{Naive Dynamic Analysis}
\label{sec:naive-dyanamic}
Our dynamic analysis tool uses Playwright’s async Python API and the Chrome DevTools Protocol (CDP) to monitor fingerprinting without modifying the original script. By evaluating targets in the page context and setting \texttt{Debugger.setBreakpointOnFunctionCall} hooks on relevant APIs (navigator, screen, canvas, audio), the tool maps IDs to labels. When an API is accessed, CDP emits a \texttt{Debugger.paused} event; the handler records the label and immediately resumes execution, minimizing timing disruptions.

Because it avoids code rewriting or monkey-patching, the tool preserves execution semantics and evades basic detection. While stability was occasionally impacted by CDP timing conflicts during page load, this minimally invasive, engine-level monitoring successfully revealed runtime fingerprinting activity post-obfuscation, confirming that dynamic defenses remain robust against WASM-based structural obfuscation.

\bibliographystyle{splncs04}
\bibliography{main}

\end{document}